\documentclass[%
 reprint,
superscriptaddress,
nofootinbib,
 amsmath,amssymb,
 aps,
 prl,
]{revtex4-2}

\usepackage{graphicx}
\usepackage{dcolumn}
\usepackage{bm}

\begin{document}

\graphicspath{{figures/}}

\preprint{APS/123-QED}

\title{Superlinear Precision and Memory in Simple Population Codes}

\author{Jimmy H. J. Kim}
\affiliation{
 Department of Physics and Astronomy, Northwestern University, Evanston, Illinois, 60208, USA
}
\author{Ila Fiete}
\affiliation{
 McGovern Institute for Brain Research, Massachusetts Institute of Technology, Cambridge, Massachusetts, 02139, USA
}
\author{David J. Schwab}
\affiliation{
 The Graduate Center, City University of New York, New York, New York, 10016, USA
}

\date{\today}

\begin{abstract}
The brain constructs population codes to represent stimuli through widely distributed patterns of activity across neurons. An important figure of merit of population codes is how much information about the original stimulus can be decoded from them. Fisher information is widely used to quantify coding precision and specify optimal codes, because of its relationship to mean squared error (MSE) under certain assumptions. When neural firing is sparse, however, optimizing Fisher information can result in codes that are highly sub-optimal in terms of MSE. We find that this discrepancy arises from the non-local component of error not accounted for by the Fisher information. Using this insight, we construct optimal population codes by directly minimizing the MSE. We study the scaling properties of MSE with coding parameters, focusing on the tuning curve width. We find that the optimal tuning curve width for coding no longer scales as the inverse population size, and the quadratic scaling of precision with system size predicted by Fisher information alone no longer holds. However, superlinearity is still preserved with only a logarithmic slowdown. We derive analogous results for networks storing the memory of a stimulus through continuous attractor dynamics, and show that similar scaling properties optimize memory and representation.
\end{abstract}

\maketitle


Information about sensory stimuli or motor variables is often encoded in the joint activity of large populations of neurons. In a classic form of such population coding, neurons fire selectively with a ``bump'' of elevated activity around a certain preferred value of the encoded variable. Such bump codes, because of their ubiquity in the brain \cite{Georgopoulos:1982, Lee:1988, Taube:1998, Wimmer:2014, Kim:2017} and amenability to quantitative analysis, have been the subject of intense theoretical scrutiny \cite{Georgopoulos:1986, Salinas:1994, Snippe:1996, Oram:1998, Pouget:2000, Sompolinsky:2001}.

\begin{figure}
	\centering
	\includegraphics[width=\linewidth]{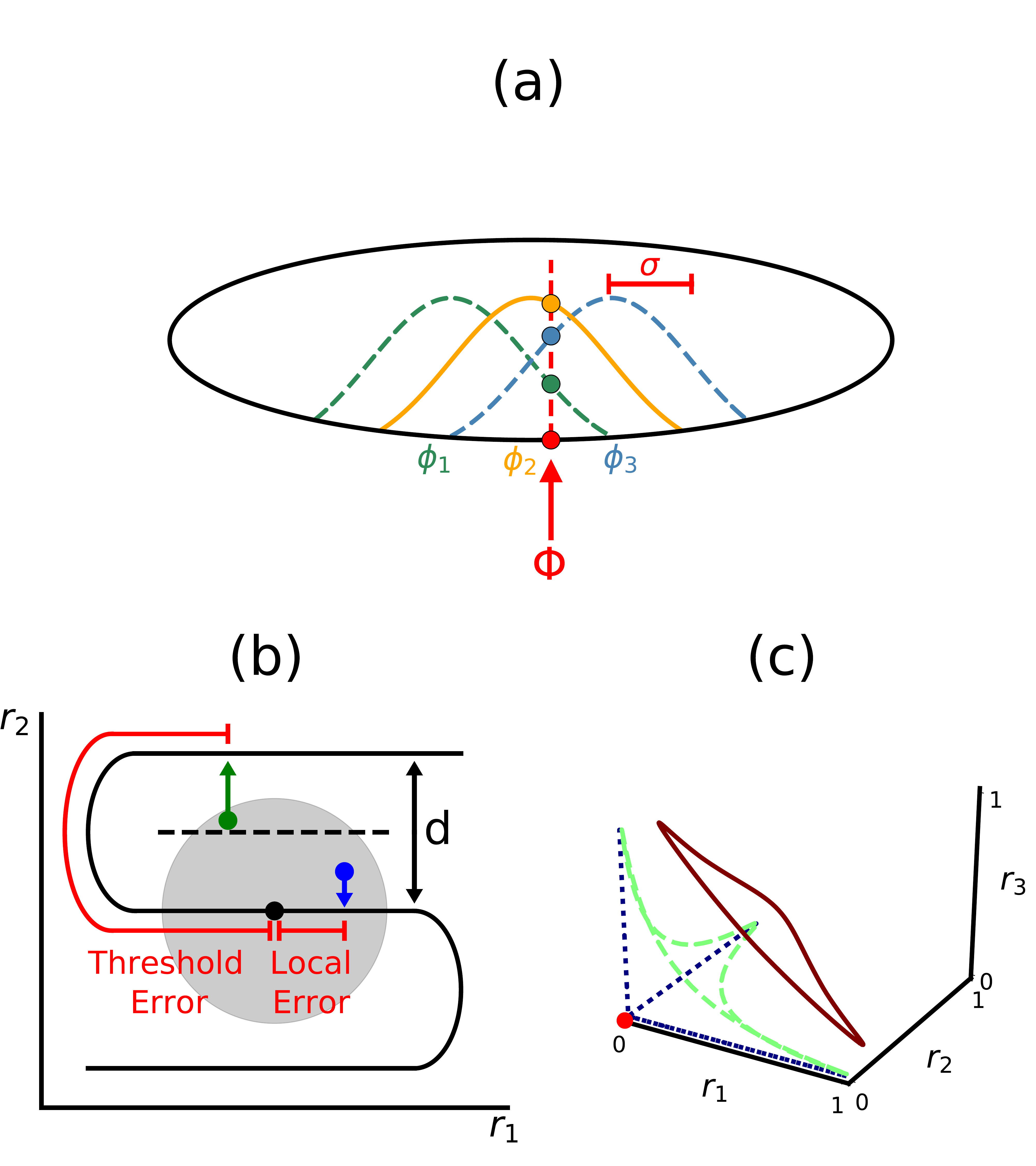}
	\caption{(a) Schematic of a bump code. Ellipse represents a ring of neurons. Tuning curves of three adjacent neurons are shown, centered at preferred stimulus values $\phi_i$. $\Phi$ is the input stimulus value that causes the neurons to fire with firing rate given by the tuning curves (dots). $\sigma$ is the tuning curve width, common to all neurons. (b) Cartoon of an abstract coding space. The curve traces out the deterministic coding line within the activity state space $\vec{r}$. A particular stimulus value corresponds to a particular point on the coding line (black dot). Due to noisy neural responses, the actual response may occur anywhere in the noise ball (gray). Ideally, a response will get mapped to a nearby point on the coding line (lower arrow). Beyond a certain threshold (dashed line; $d/2$ away from the black dot) however, the response will get mapped to a far-away point (upper arrow), causing a large threshold error. (c) Coding lines embedded in 3-dimensional activity space for $\sigma=1.0$ (solid), $0.66$ (dashed), and $0.1$ (dotted). As the tuning width decreases, the length of the coding line increases. However, the coding line also approaches the origin (dot), where no neurons respond, resulting in a threshold error. The dots correspond to uniformly distributed angles.}
	\label{fig1}
\end{figure}

Fig. \ref{fig1}a shows a schematic of a bump code, where a network of $N$ neurons encodes a one-dimensional periodic stimulus parametrized by angle $\phi \in [0,2\pi]$. When the stimulus $\phi$ is presented, each neuron $i$ independently fires Poisson spikes at a rate $\lambda_i$:
\begin{equation}\label{eq:tc}
    \lambda_i = \lambda_{max}\left[(1-\epsilon)f_i(\sigma,|\phi-\phi_i|)+\epsilon\right] 
\end{equation}
Here, $\lambda_{max}$ is the peak firing rate and $f_i(\sigma, x)$ is a unimodal function with width $\sigma$ and the peak value of 1 at $x=0$. We set the baseline firing rate $\epsilon=0$, but later find that this assumption can be relaxed. The preferred angles $\phi_i$'s of the neurons are evenly distributed across the stimulus domain, such that $\phi_i = 2\pi i/N$. Equation \eqref{eq:tc}, which maps each stimulus value to the expected response of a neuron, is known as the neuronal \emph{tuning curve}. Such unimodal tuning curves are widely observed in sensory and motor peripheries and even in cognitive areas \cite{Schwartz:1988, Miller:1991, Young:1992}. If the spikes are collected for time $T$ while the stimulus $\phi$ remains present, the number of spikes fired by each neuron will be distributed as $r_i \sim Poiss(T\lambda_i)$. The population response $\vec{r}=(r_1, \dots, r_N)$ constitutes the neural encoding of stimulus $\phi$.

A natural measure of encoding performance in a population code is the mean squared-error (MSE) in estimating the stimulus from the populaton response. In general, MSE is difficult to compute thus a common approach has been to instead compute the Fisher information (FI) \cite{Paradiso:1988, Seung:1993, Abbott:1999, Harper:2004, Toyoizumi:2006, Yarrow:2012}:
\begin{equation}
J(\phi) = \Big<\big(\frac{\partial}{\partial\phi}\log p(\vec{r}|\phi)\big)^2\Big>_{p(\vec{r}|\phi)}
\end{equation}

For any unbiased estimator $\hat{\phi}$, the MSE is bounded from below by the inverse of FI via the Cram\'{e}r-Rao bound. \cite{Rao:1945, Cramer:1946}:
\begin{equation}
    \text{var}(\hat{\phi}) \geq 1/{J(\phi)}
\end{equation}
Averaging over all possible stimulus values, the quantity $\left<1/J(\phi)\right>_{p(\phi)}$ is a lower bound on overall MSE.

In general, a bigger population will enable more accurate decoding. However, neurons are costly to maintain so the brain may optimize other coding parameters. A parameter of particular interest is the tuning curve width $\sigma$. For bump codes representing a scalar variable, FI grows as $N/\sigma$ \cite{Seung:1993,Zhang:1999, Dayan:2001}. Thus, if $\sigma$ remained constant, FI would scale linearly with $N$.

While it may appear that we can achieve infinite precision by sending $\sigma\to0$, this is not the case because the relevant quantity, $\left<1/J(\phi)\right>_{p(\phi)}$, diverges from $1/\left<J(\phi)\right>_{p(\phi)}$ in the regime in which the tuning curves are too narrow to span the space between each neuron's bump center (i.e., the support of the tuning curves does not cover the stimulus domain). The correct FI-optimal $\sigma$ that minimizes the former actually scales as $1/N$ \cite{Berens:2011}, and so the optimal FI scaling of precision is superlinear, scaling as $N^2$.

Unfortunately, the Cram\'{e}r-Rao bound is guaranteed to be ``tight'' only when the number of samples collected for the estimate tends to infinity. When the number of spikes obtained from the population is small, the inverse FI can severely underestimate the true MSE, as demonstrated in \cite{Bethge:2002, Yaeli:2010, Berens:2011}. This finding calls into question whether any superlinear scaling of MSE is actually achievable in bump codes. Thus, we are left with two open questions which we answer in this Letter: What is the MSE-optimal scaling of tuning width in classical bump codes? And, is superlinear coding possible?

A natural way to start is to ask when the inverse FI and MSE become decoupled. For this, we adopt a geometric view of coding. A redundant code of an analog variable may be viewed as an embedding of a lower-dimensional manifold (encoded variable) into a higher-dimensional space (coding variables, i.e. neural response). Thus, our bump population code for a scalar variable is an embedding of a line into the $N$-dimensional activity state space $\vec{r}$. In Fig. \ref{fig1}b, the solid black coding line corresponds to the noiseless neural responses $T\vec{\lambda}$ as a parametric function of the encoded variable $\phi$. Noise in the neural responses perturbs the network state away from the coding line, and a decoder must map the perturbed state back onto it. When the noise is small, a good decoder can map the state back to the vicinity of the original point on the coding line; the small remaining errors are \emph{local}. When the noise magnitude exceeds a threshold value, the perturbed state, and thus its reconstructed estimate, is closer to a distant point on the coding line. Such errors are called \emph{threshold} errors \cite{Yoo:2016}. Both types of error contribute to the MSE, but FI only takes into account of local errors. Thus, when the threshold errors proliferate, the MSE grows and parts ways from the inverse FI.

The tuning curve width affects the layout, length, and spacing of the coding line, and through them the probability of threshold errors. If the total volume of state space is held fixed (equivalent to fixing the minimum and maximum firing rate of neurons and the number of neurons) as the tuning curve is narrowed, the coding line increases in length, resulting in a smaller local error as a fraction of the range of the variable. However, the longer coding line is packed more closely near the axes of the space and near the origin (Fig. \ref{fig1}c). As we see below, this increases threshold error probability.

We now use the above insight to heuristically derive a simple analytic expression for how MSE scales with network size in bump population codes. The total MSE can be written as a sum of two terms, arising from local and threshold errors:
\begin{equation}
    MSE \simeq E_{local}(1-p_{th}) + E_{th}p_{th}
\end{equation}
where $E_{local}$ ($E_{th}$) corresponds to the expected value of local (threshold) squared errors and $p_{th}$ is the probability of threshold errors.

As noted above, for unimodal tuning curves, the FI for a one-dimensional stimulus has the following scaling:
\begin{equation}
    J \simeq \alpha N/\sigma \label{eqJ}
\end{equation}
where $\alpha \equiv \alpha(T,\lambda_{max})$ is a prefactor dependent on $T$ and $\lambda_{max}$, which is fixed. We assume that $E_{local}$ is accurately described by the Cram\'{e}r-Rao bound and that the regime of interest satisfies $\sigma > 1/N$. We will subsequently verify the consistency of this assumption.

We next consider the threshold error term. If at any stimulus value few neurons respond and do so with small rates, there is a finite probability that no spikes will be fired. In such a trial, the decoder must guess an angle from no data, resulting in a large error of $\mathcal{O}(1)$. This is the manifestation of threshold error in our system, and corresponds to the intersection of a noise ball with the origin in the geometric view of Fig. \ref{fig1}c. To reduce such error, the neurons must code redundantly in the sense of multiple neurons covering the same angular space to ensure that at least some will respond for any stimulus. It is thus expected that the optimal width will decrease more slowly than $\sim1/N$ to ensure that the threshold errors remain comparable to the inevitable local errors.

The equation for the MSE now becomes:
\begin{align}
    MSE & \simeq \frac{\sigma}{\alpha N}(1-p_0) + \beta p_0 \label{eqMSE}
\end{align}
where $p_0$ is the probability that no spikes are fired by any neurons during the observation time $T$ and $\beta = \pi^2/3$ is the mean-squared error in the case of random guessing over the circle. For broad enough tuning curves, the total firing rate in the population is nearly independent of stimulus value and is given by $\lambda_{pop} \simeq \gamma \lambda_{max} N \sigma$, where $\gamma$ is a constant determined by the tuning curve shape, e.g. $1/\sqrt{2\pi}$ for Gaussian \cite{Yaeli:2010}. Thus, we have $p_0\simeq e^{-T\lambda_{pop}} = e^{-\gamma T\lambda_{max}N\sigma}$. Above, we made the assumption that $\sigma$ decreases more slowly than $1/N$. This implies both that fluctuations in $\lambda_{pop}$ is negligible and $p_0 \ll 1$. The scaling of optimal width $\sigma^*$ that we presently derive is consistent with this assumption.

From Equation \eqref{eqMSE}, the value of $\sigma$ that minimizes MSE is readily computed to be
\begin{align}
    \sigma^*(N) & \simeq \frac{\ln(\alpha\beta \gamma T \lambda_{max} N^2)}{\gamma T\lambda_{max}N}\\
    & \sim \frac{2}{\gamma T\lambda_{max}}\frac{\ln N}{N}, \text{as }N\to\infty \label{eqsigopt}
\end{align}

With this form for $\sigma$, the optimal MSE then scales as
\begin{equation}
    MSE^* \sim \frac{\ln(\alpha\beta\gamma T\lambda_{max} N^2)}{\alpha\gamma T\lambda_{max}N^2} + \frac{1}{\alpha\gamma T\lambda_{max} N^2}
\end{equation}

Due to $\ln N$ in the numerator, the first term is dominant for large $N$ and scales as $\frac{\ln N}{N^2}$. This provides the optimal scaling of MSE for neurons with unimodal tuning curves encoding a one-dimensional stimulus. An important note is that since the first term corresponds to the local error, the Cram\'{e}r-Rao bound is in fact asymptotically tight with the optimal scaling in Equation \eqref{eqsigopt}. This furthermore implies that the extra error, i.e. the absolute difference between $MSE(\sigma^*)$ and $1/J(\sigma^*)$, the inverse FI at the optimal tuning width, should scale as $1/N^2$.

\begin{figure}
	\centering
	\includegraphics[width=\linewidth]{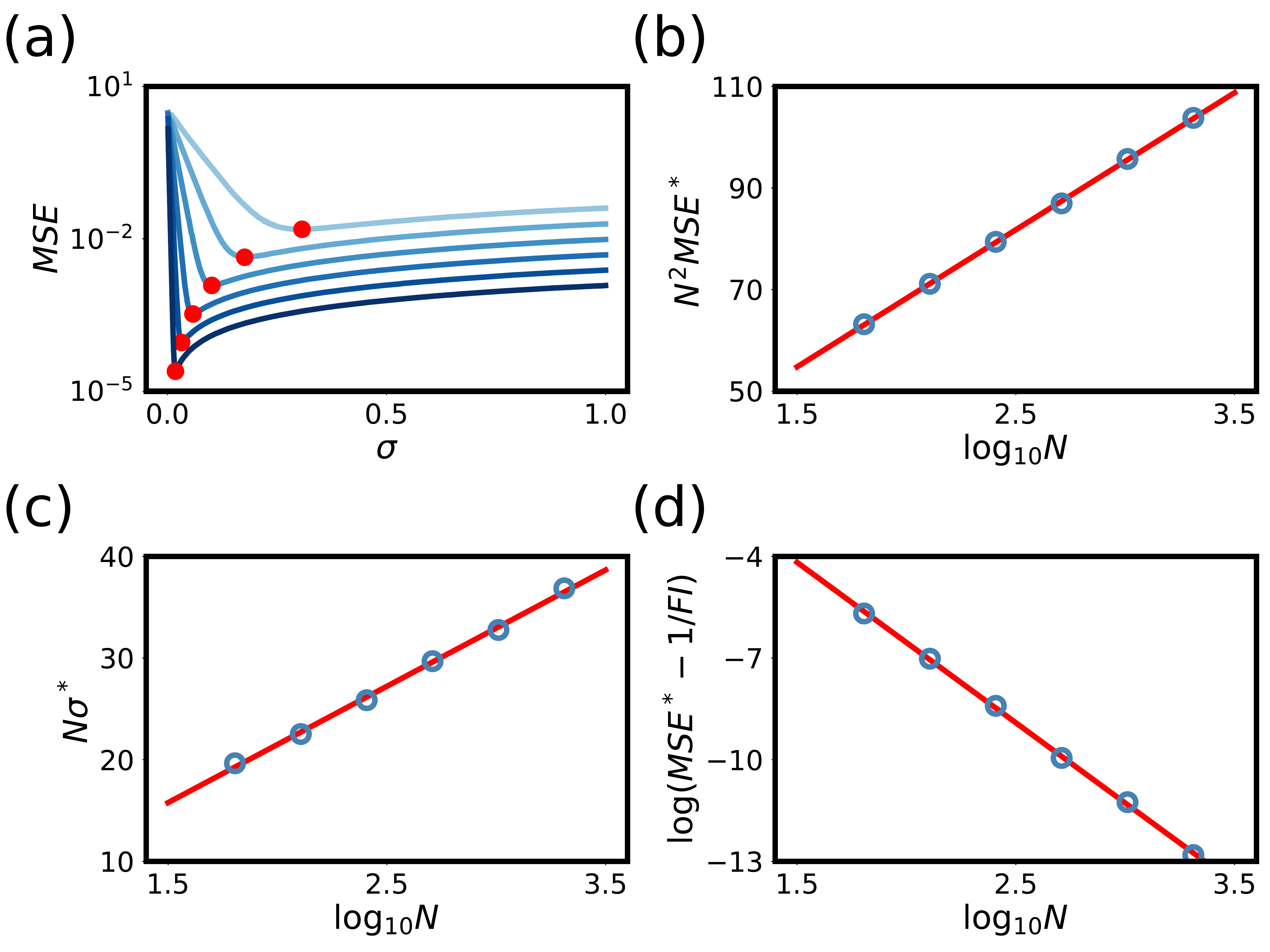}
	\caption{(a) MSE versus tuning curve width obtained from simulations with maximum likelihood decoding for system sizes $N = 64, 128, 256, 512, 1024, 2048$ (top to bottom). Dots indicate the minima ($MSE^*$). (b) Plot of $N^2MSE^*$ versus $\log N$. (c) Plot of $N\sigma^*$ versus $\log N$. (d) Plot of $\log(MSE^*-1/FI)$ versus $\log N$.}
	\label{fig2}
\end{figure}

To test these predictions, we performed simulations of the encoding/decoding process that employed maximum likelihood decoding of neurons with Gaussian tuning curves. We chose various system sizes and measured MSE as a function of tuning curve width for each value of $N$ (Fig. \ref{fig2}a). To check the consistency with our predictions, we rescaled the minimal MSE for each $N$ and plotted $N^2MSE^*$ versus $\log N$ (Fig. \ref{fig2}b), which should be linear if the predicted scaling is correct. We similarly rescaled the optimal tuning curve width and plotted $N\sigma^*$ versus $\log N$ (Fig. \ref{fig2}c), which should also be linear. We finally plotted the extra threshold error beyond the local error (Fig. \ref{fig2}d). In all cases, the predicted scaling was observed. These numerical results together provide strong evidence that our simple estimate captures the correct asymptotic behavior and that indeed $1/MSE$ can scale nearly as $N^2$, with only a logarithmic slowdown. Moreover, the optimal tuning curve width scales as $\log(N)/N$, an adjustment from the Fisher information-derived  $1/N$ scaling. We emphasize that this is a general result for any unimodal symmetric tuning curves as long as neurons are homogeneous, densely distributed, and fire independent Poisson spikes with zero baseline.

We decided to test whether the result still holds if the neurons are assumed to have a small (compared to the maximum) nonzero baseline firing rate. The above derivation cannot be reused because the main cause of threshold errors is now no longer non-response of the stimulus-driven neurons but higher responses by neurons far from the stimulus. Nevertheless, a heuristic derivation suggests that the basic scaling relations still remain the same, and numerical results also corroborate this. Details can be found in the Supplemental Material.

\begin{figure}
	\centering
	\includegraphics[width=\linewidth]{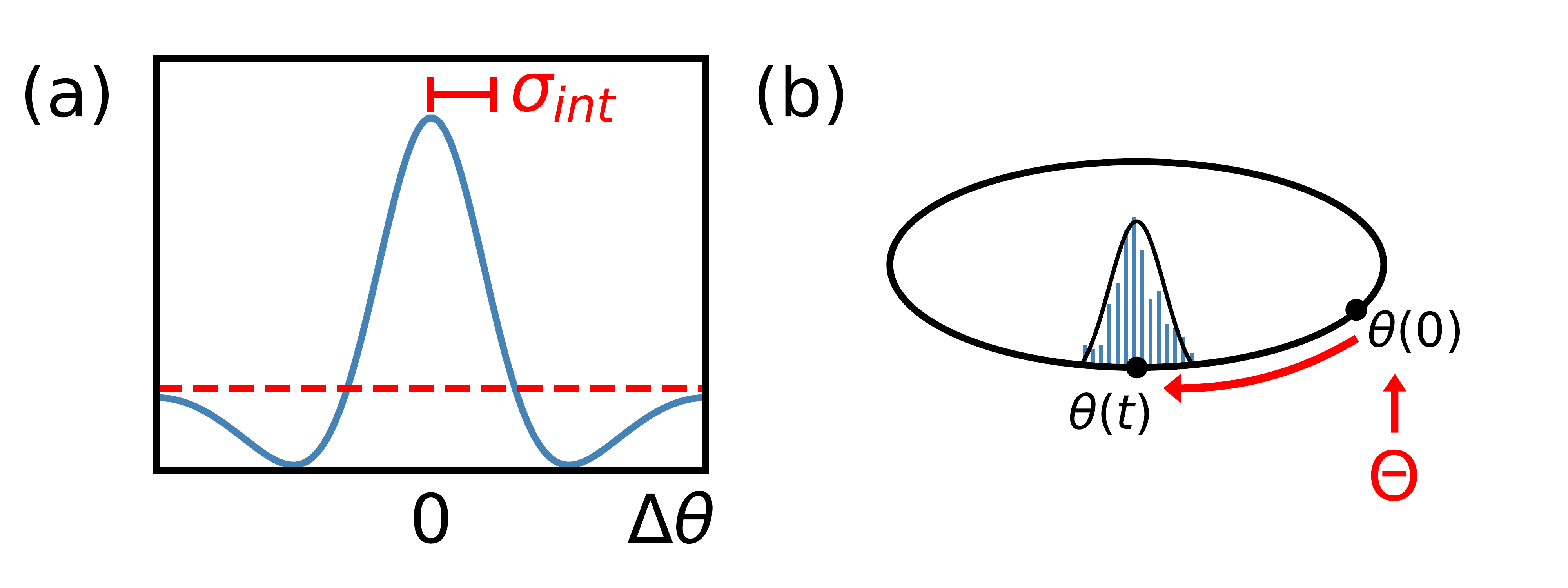}
	\caption{(a) Interaction matrix between neurons as a function of preferred angle difference. The dashed line distinguishes excitatory interaction weights above and inhibitory interaction weights below. $\sigma_{int}$ determines the range of excitatory interaction and is linearly related to the bump width, which is analogous to the tuning curve width of the sensory network. (b) Schematic of the memory network. Ellipse represents a ring of neurons. The bump of local activity (bars) was originally initialized at $\theta(0) = \Theta$ by an external input but subsequently diffused to $\theta(t)$.}
	\label{fig3}
\end{figure}

The brain also uses population representations to store short-term memory of continuous variables, which can be maintained as the dynamical attractor states of a continuous attractor network \cite{Seung:1996, Camperi:1998, Compte:2000}. 
The performance of such memory networks is also known to be bounded by FI \cite{Burak:2012}. This motivated us to ask whether the considerations discussed above may also apply in a memory setting. For direct comparison, we consider neurons storing a periodic one-dimensional variable, through bump-like activity profiles. To achieve a persistent bump in the absence of a stimulus, the neurons are arranged in a ring and each neuron interacts with others via short range excitatory and long range inhibitory connections (Fig. \ref{fig3}a). The excitatory interaction width $\sigma_{int}$ plays an analogous role to the tuning curve width we considered earlier. The result of this connectivity structure is that a persistent bump of local activity is dynamically maintained around an initial location determined by the external input stimulus (which is subsequently taken away) \cite{Skaggs:1995, Zhang:1996}. Due to neural noise, the bump location does not remain stationary, but rather performs diffusive motion away from its initial location \cite{Burak:2012} (Fig. \ref{fig3}b). This diffusion amounts to erasure of the stored information (initial stimulus location), and a desirable memory network would have lower diffusivity.

\begin{figure}
	\centering
	\includegraphics[width=\linewidth]{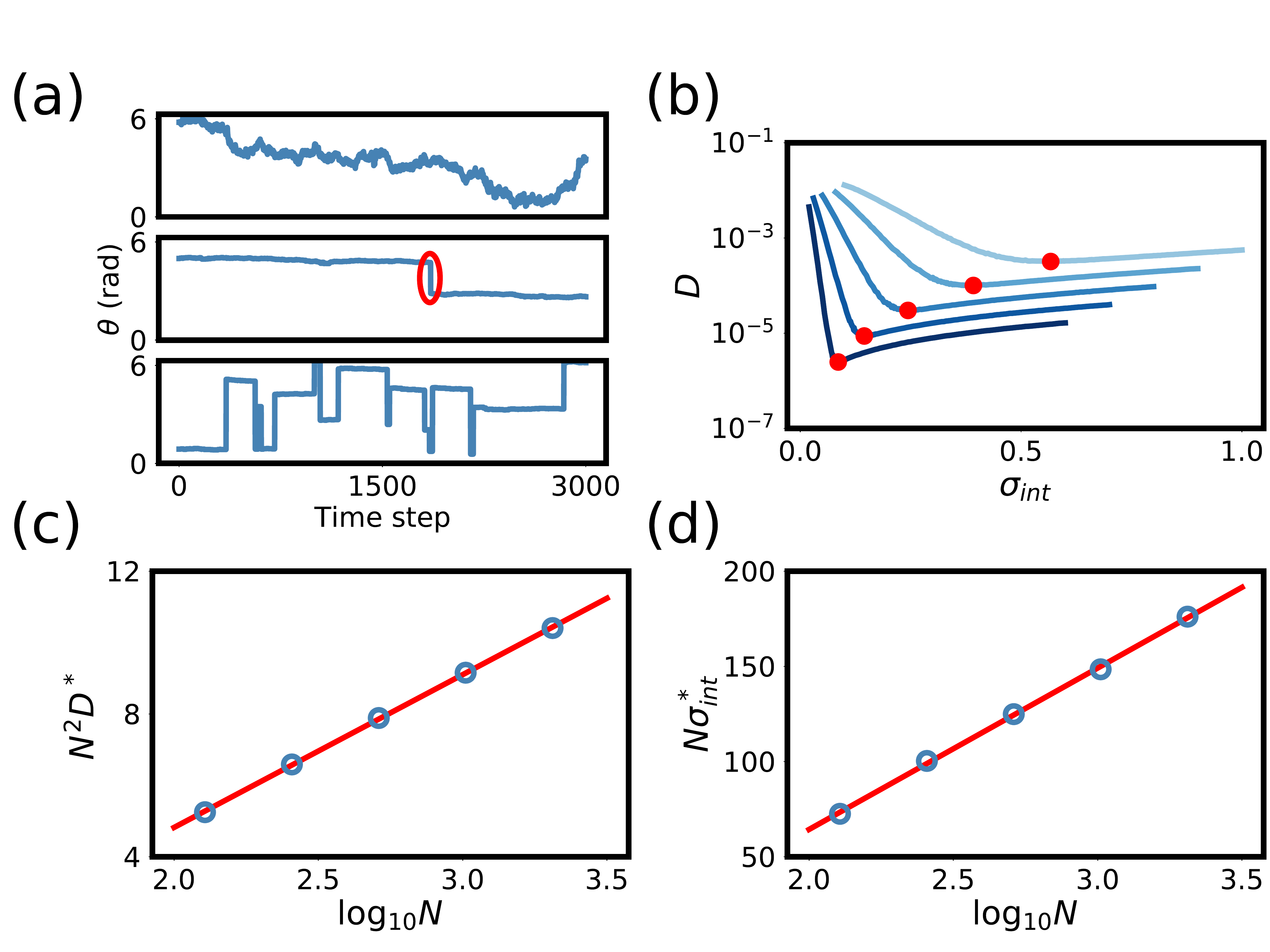}
	\caption{(a) Example traces of bump dynamics for broad to narrow (top to bottom) interaction widths. We tracked and plotted the location of the bump peak. For broad interactions, bump motion is diffusive. For narrower widths, bump motion exhibits epochs of diffusive dynamics interrupted by large jumps (one indicated by an ellipse). It can be inferred that finding optimal performance depends on balancing diffusive dynamics with jump probability. (b) Effective diffusion constant, measured by computing the slope of the mean squared displacement of bump location over short time intervals, as a function of interaction width, $\sigma_{int}$, for system sizes $N = 128,256,512,1024,2048$ (top to bottom). Dots indicate the minima. (c) Plot of $N^2D^*$ versus $\log N$. The straight line is consistent with the scaling of minimal mean-squared error found for the sensory system. (d) Plot of $N\sigma_{int}^*$ versus $\log N$, where $\sigma_{int}^*$ is the interaction width that minimizes bump diffusivity.}
	\label{fig4}
\end{figure}

In fact, we find that the motion is diffusive only for large enough interaction width. For smaller interaction width, diffusive motion is interrupted by large jumps where the activity bump stochastically shrinks and spontaneously reassembles in a potentially distant location (Fig. \ref{fig4}a). These non-local jumps can be understood as the dynamical analogue of threshold errors in the sensory network. Clearly, such large jumps are catastrophic for memory performance. Thus a similar trade-off exists for the memory network: decreasing the interaction width tightens the bump and decreases its diffusivity, but it also increases the chance of complete destabilization and non-local reformation of the bump.

We performed dynamical simulations for networks of various sizes and varied $\sigma_{int}$. For each network, we computed the diffusion constant as the slope of mean squared displacement versus time for short durations, averaged over the entire simulation. Although the motion was not always purely diffusive, we can nonetheless extract an effective diffusion constant in this fashion, and mean squared displacement was linear over short times for all networks. From the results (Fig. \ref{fig4}b), we found the minimum diffusivity and the corresponding optimal interaction width and plotted $N\sigma^*_{int}$ and $N^2D^*$ versus $\log N$ (Fig. \ref{fig4}c-\ref{fig4}d). In both cases the scaling is linear, indicating that the optimal memory network exhibits the same $N$ scaling as we found for the sensory system.

Thus, it is found that the decomposition of error into local and non-local terms is a phenomenon common to both population coding and short-term memory. Fisher information, being a partial derivative, is fundamentally a local quantity and can only take into account of the first term. Neglecting the existence of the non-local source of error inevitably leads to failure of any FI-based approach to characterizing and optimizing model performance. Conversely, we hope to have demonstrated that FI can still be used fruitfully if it is complemented with an appropriate characterization of the remaining error.


%

\end{document}